%% file: NuMIatBooNE_web_v2.tex
\documentclass[aps,prl,twocolumn,groupedaddress,showpacs,preprintnumbers,superscriptaddress]{revtex4}

\usepackage{graphicx}
\usepackage{dcolumn}
\usepackage{bm}
\usepackage{array,hhline,dcolumn} 
\usepackage{epsfig}
\usepackage{epstopdf}
\begin{document}  
\title{First Measurement of $\nu_\mu$ and $\nu_e$ Events in an Off-Axis Horn-Focused Neutrino Beam}

\input{collaborators_v2}

\date{\today}
\begin{abstract}
We report the first observation of off-axis neutrino interactions in the MiniBooNE detector from the NuMI 
beamline at Fermilab. The MiniBooNE detector is located 745~m from the NuMI production target,
at 110~mrad angle ($6.3^{\circ}$)  with respect to the NuMI beam axis. Samples of charged current quasi-elastic $\nu_{\mu}$ and $\nu_e$ interactions are analyzed and found to be in agreement with expectation. This provides a direct verification of the expected pion and kaon contributions to the neutrino flux and validates the modeling of the NuMI off-axis beam. 
\end{abstract}


\maketitle
Conventional neutrino beams from high-energy proton accelerators serve as important tools for studying neutrino 
characteristics and the fundamental properties of matter involving interactions of neutrinos. 
Such beams typically arise from the two-body decays of $\pi$ and $K$ mesons 
produced by a proton beam impinging upon a nuclear target.
The mesons leave the target with a significant angular divergence. 
The flux of neutrinos in such a wide band beam at distance $d$ from the meson decay point and at an angle $\theta$ with 
respect to the parent meson direction is given by
 \begin{equation}
 \Phi_{\nu} \approx  \frac{1}{4 \pi d^2} \Big(\frac{2 \gamma}{1 + \gamma^2 \theta^2} \Big)^2,
\end{equation}
where $\gamma$ is the Lorentz boost factor of mesons~\cite{kopp_phys_rep}.
To obtain a more intense neutrino flux, it is essential to
focus the mesons produced in the target. To accomplish this, neutrino experiments such as MiniBooNE~\cite{MB_osc} 
and MINOS~\cite{adamson} use focusing magnetic horns to 
direct the mesons toward downstream detectors.
The energy of $\nu_{\mu}$s from two-body decays is given by
\begin{equation}
E_{\nu}\approx{\frac{{\Big(1-{\frac{{m_{\mu}^{2}}}{{m_{\pi,K}^{2}}}
}\Big)E_{\pi,K}}}{{1+\gamma^{2} \tan^2\theta}}},
\end{equation}
where $m_{\pi,K}$ ($E_{\pi,K}$) is the mass (energy) of the $\pi$, $K$ parent, and  $m_\mu$ is the muon mass.
Brookhaven experiment E889 proposed~\cite{bnl} an off-axis beam because, at a
suitable off-axis angle $\theta$, the neutrino flux is confined to a relatively narrow band of energies,
which is useful in limiting backgrounds in searches for the oscillation transition $\nu_{\mu} \rightarrow \nu_e$. 
Future neutrino oscillation searches by the T2K~\cite{T2K} and NO$\nu$A~\cite{Nova}
experiments plan to use off-axis horn-focused beams. 

\begin{figure}[ptb]
\includegraphics[trim=10mm 73mm 90mm 73mm, height=4.8cm, width=4.95cm, angle =-90]{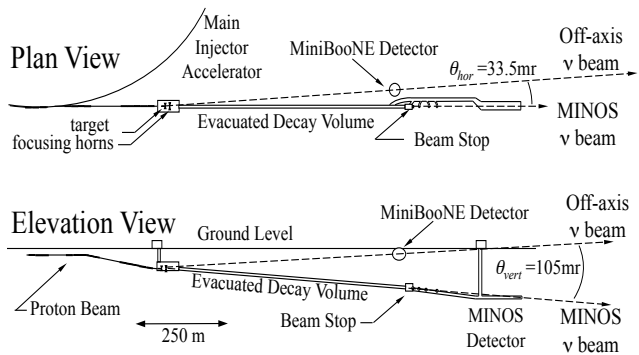}
\caption{\label{fig:prl_fig1} Plan and elevation views of the NuMI beamline with respect to the MiniBooNE detector. 
The MiniBooNE detector is located 745~m from the NuMI production target,
at 110~mrad ($6.3^{\circ}$)  with respect to the NuMI beam axis. The length of the NuMI target hall is 45~m;
the length of the evacuated decay volume is 675~m, and the distance from the target to the Minos Near Detector
is 1040~m.}
\end{figure}

The MiniBooNE detector, located at an angle of 110~mrad ($6.3^{\circ}$) with respect to the NuMI beam axis  (see Fig.~\ref{fig:prl_fig1}),
provides a unique opportunity to perform the first measurement of neutrino interactions from an off-axis horn-focused beam. 
In addition to demonstrating the off-axis beam concept, the measurement verifies
the predicted fluxes from $\pi/K$ parents in the NuMI beam, and probes the off-axis intrinsic $\nu_e$ contamination,
required for future $\nu_\mu \rightarrow \nu_e$ appearance searches.

The NuMI beam points toward the
MINOS Far Detector, located in the Soudan Laboratory in Minnesota.
Neutrinos are produced by 120~GeV protons incident on a carbon target.
In the period studied here, the beam intensity was up to $3\times10^{13}$ 
protons on target per spill at a typical repetition rate of 0.48 Hz. 
Positive $\pi$ and $K$ mesons produced in the target are focused down the decay pipe using two
magnetic horns.
Neutrinos from two-body decays of pions are more forward directed than those from kaons
due to the difference in rest mass of the decaying mesons.
As a result, the off-axis component coming from pions is suppressed relative 
to the kaon component. Decay in flight of poorly focused pions can only occur close to the NuMI target, since they are 
stopped by shielding around the target area.
The NuMI beam also provides a large sample of $\nu_{e}$ events
in the MiniBooNE detector. The $\nu_{e}$'s result primarily from the three-body decay 
of kaons and thus have a wider range of energies.
The stability of the neutrino beam is monitored using the muon monitors at the
end of the decay pipe and the MINOS Near Detector. The direction of the
neutrino beam, its intensity, and its energy spectrum were all found to 
be very stable over the data taking period~\cite{adamson}.
\begin{figure}[ptb]
\includegraphics[height=4.1cm, width=8.6cm]{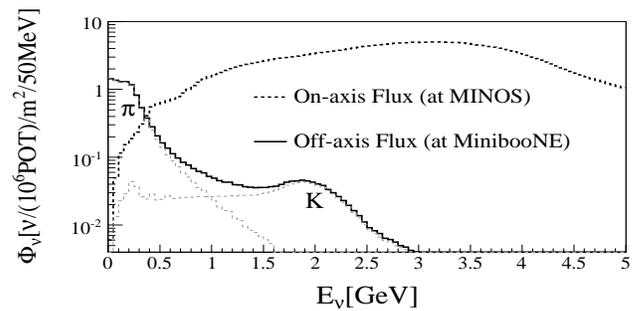}
\caption{
Comparison of the predicted NuMI off-axis and on-axis fluxes including all neutrino species.
The off-axis flux is separated into contributions from charged $\pi$ and $K$ parents. 
}
\label{fig:prl_fig_flux}
\end{figure}

Detailed  {\sc GEANT3}-based~\cite{geant3} Monte Carlo (MC) simulations of the beam, 
including secondary particle production,
 particle focusing, and transport, are performed to calculate the flux as a
function of neutrino flavor and energy.  
The yield of pions and kaons from the NuMI
target is calculated using the {\sc FLUKA} cascade model~\cite{fluka}. 
The beam modeling includes downstream interactions in material other than the target that produce 
hadrons decaying to neutrinos.
These interactions are modeled using a {\sc GEANT3} simulation, 
configured to use either {\sc GFLUKA}~\cite{geant3} or {\sc GCALOR}~\cite{gcalor} cascade models. 
The NuMI neutrino flux at the MiniBooNE detector is shown in Fig.~\ref{fig:prl_fig_flux}.
Pions (kaons) produce neutrinos with average energies of about 0.25~GeV (2~GeV).

These neutrinos are detected in the MiniBooNE detector~\cite{MB_detector} which is a 12.2~m 
spherical tank filled with 800 tons of pure mineral oil. The detector triggers on every NuMI beam spill and 
the detector activity is recorded in a 19.2~$\mu s$ window beginning about 1~$\mu s$ before the start of 
a $\sim$10~$\mu s$ wide spill. 
The stability of the NuMI beam was confirmed by studying the rate of neutrino events in the MiniBooNE detector as a function of time.
The time and charge of photomultiplier-tubes (PMT) in the detector are used 
to reconstruct the interaction point, event time, energies, and particle tracks resulting from neutrino interactions. 
Neutrino interactions in the detector are simulated with the {\sc NUANCE} event generator package~\cite{Nuance}, 
with modifications to the quasi-elastic (QE) cross-section as described in~\cite{MB_ccqe}. Particles generated 
by {\sc NUANCE} are propagated through the detector, using a {\sc GEANT3}-based simulation which describes 
the emission of optical and near-UV photons via Cherenkov radiation and scintillation. Neutrino induced events are 
identified by requiring the event to occur during the NuMI beam spill, after rejection of cosmic ray muons and 
muon decay electrons~\cite{MB_osc}. For the sample satisfying these selection criteria, NuMI neutrinos are predicted to interact via 
charged-current (CC) QE scattering (39\%), CC single pion production (31\%), neutral current (NC) single pion 
production (14\%), multi-pion production (9\%), deep inelastic scattering (4\%), and other interactions (3\%). 
The predicted event composition is $\nu_{\mu}$ : $\bar{\nu}_{\mu}$ : $\nu_{e}$ : $\bar{\nu}_{e}$ $\sim$ 0.81 : 0.13 : 0.05 : 0.01.

The data set analyzed here corresponds to $1.42\times10^{20}$ protons delivered to the NuMI target 
from June 22, 2005 to March 2, 2007.
The MC in all cases has been normalized to this number of protons. 
There is a 2\%  uncertainty in the number of protons on target.
Neutrino interactions are identified with the likelihood-based algorithm used in~\cite{MB_osc}.

The high rate and simple topology of  $\nu_{\mu}$ CCQE events
provides a useful sample for understanding
the $\nu_{\mu}$ spectrum and verifying the MC prediction for 
$\nu_{e}$ production. The identification of $\nu_{\mu}$
CCQE events is based upon the detection of the primary stopping muon and the associated
decay electron as two distinct time-related clusters of PMT hits, called 'subevents':
$\nu_{\mu}+n\rightarrow\mu^{-}+p,\;\;\;\mu^{-}\rightarrow e^{-}+\nu_{\mu}+\bar{\nu}_{e}$.
We require the first subevent to have a reconstructed position within 5~m of the
center of the detector. 
The decay electron requirement substantially reduces CC single
$\pi^{+}$ contamination because most CC $\pi^{+}$ events contain a second decay
electron from the $\pi^{+}$ to $\mu$ decay chain. 
Additional rejection of non-$\nu_{\mu}$ CCQE events in the sample is achieved 
by a requirement on the relative $\mu$ and e likelihoods, maximized under a given particle hypothesis, log($L_{e}/L_{\mu})<0.02$. 
This selection criterion is 24\% efficient in selecting
$\nu_{\mu}$ CCQE candidates, resulting in a 69\% pure $\nu_{\mu}$ CCQE
sample. The most significant background contribution to the $\nu_{\mu}$ CCQE sample results from 
CC single $\pi^{+}$ production (78\%) where the $\pi^{+}$ is undetected. 
A total of 17659 data events pass this $\nu_{\mu}$ CCQE selection criteria,
compared to the prediction of 18247$\pm$3189 in the 0.2 $< E_\nu < $ 3.0 GeV range; 
the uncertainty includes systematic errors
associated with the neutrino flux, neutrino cross-sections,
and detector modeling.

The flux uncertainties include particle production in the NuMI target, modeling of the 
downstream interactions, and kaons stopped in the NuMI beam dump.
The flux uncertainty also includes the uncertainty arising from possible misalignment of the target, the
focusing horns and the shielding blocks although this was found to have a small effect on 
the off-axis neutrino flux~\cite{zwaska,pavlovic}. The $\pi/K$ yields off the target were tuned to match 
the observed neutrino event rates in the MINOS Near Detector~\cite{adamson}, where the same meson
decays produce significantly higher energy neutrinos.
Such tuning has a negligible effect on the off-axis beam at MiniBooNE. 
However, the difference between the tuned and the untuned $\pi/K$ yields was treated as
an additional systematic effect.  Further details of systematic uncertainties considered may be found in~\cite{adamson},
though it is important to note that the magnitudes of these systematic uncertainties due to the flux are
smaller in the off-axis case.
The cross-section uncertainties are quantified by varying the underlying
model parameters constrained by either external or
Booster neutrino beam (BNB) data. Uncertainties in the parameters describing the
optical properties of the MiniBooNE detector are constrained
by external measurements of the oil properties and by fits to calibration
samples of events in the BNB data sample~\cite{MB_osc}.

\begin{figure}[ptb]
\includegraphics[height=5.2cm, width=8.6cm]{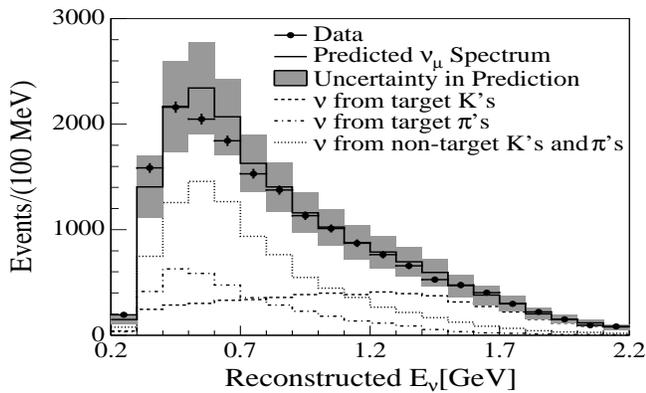}
\caption{ Reconstructed $E_{\nu}$ distribution of $\nu_{\mu}$ CCQE events.
The band indicates the total systematic uncertainty associated with the MC prediction.
The prediction is separated into contributions from kaon parents produced in the NuMI target,
pion parents produced in the NuMI target, and kaon and pion parents produced in non-target
materials. 
} 
\label{fig:prl_fig2}
\end{figure}

Reconstructed $\nu_{\mu}$ CCQE event kinematics include
the muon energy, $E_\mu$, and muon angle with respect to the neutrino beam direction, $\theta_{\mu}$. 
For both the data and MC, $\theta_{\mu}$ is approximated assuming that all
neutrinos arise from meson decays at the NuMI production target. 
In reality, mesons decay along the NuMI beamline so that the average decay distance from the target
is about 70~m. However, given the geometry of the beamline with respect to the detector  (see Fig.~\ref{fig:prl_fig1}),
such an off-axis angle change is well within the angular resolution of the detector ($\sim2^{\circ}$). 
Based on these reconstructed quantities, 
the neutrino energy, $E_{\nu}$, is calculated assuming two-body kinematics
\begin{equation}
\label{eq:EnuQE}
E_{\nu} = \frac{1}{2}\frac{2M_{p}E_{\mu}-m_{\mu}^{2}}{M_{p}-E_{\mu}
+\sqrt{(E_{\mu}^{2}-m_{\mu}^{2})}\cos\theta_{\mu}},
\end{equation}
where $M_{p}$ is the proton mass.
The $E_{\nu}$ resolution of NuMI neutrinos in MiniBooNE is $\sim$~12\% at 1~GeV.  
The $E_{\nu}$ distribution of selected $\nu_{\mu}$
CCQE events is shown in Fig.~\ref{fig:prl_fig2}, along with the MC prediction, separated
into contributions from target pions, target kaons, and non-target sources. About 50\% 
of the events in  the $\nu_{\mu}$ CCQE sample originate from parents produced in non-target materials.
Predicted pion and kaon contributions, in two energy regions, are given in Table~\ref{table:two_bins_numu}. 
\begin{table}
\caption{ Observed and predicted $\nu_\mu$-like events in two energy bins. The
predicted events are further separated into contributions from kaon and pion parents of neutrinos.}
\scriptsize
\begin{ruledtabular}
\begin{tabular}[c]{ccccc} 
$E_{\nu}$[GeV] & Data                     & MC Prediction          & $\pi$      & $K$         \\ \hline
0.2-0.9                  & 10734 $\pm$104      & 11169 $\pm$1989   & 7635      & 3534   \\
0.9-3.0                  & 6925  $\pm$    83     &  7078 $\pm$1354    & 1884      & 5194  \\ 
\end{tabular}
\end{ruledtabular}
\label{table:two_bins_numu}
\end{table}
Systematic uncertainties of the predicted event rates are given in Table~\ref{table:syst_err}.
\begin{table}[ptb]
\caption{ Systematic uncertainties of the predicted event rate in the full energy range and in 
two $E_{\nu}$ bins, for CCQE $\nu_{\mu}$ and $\nu_e$ samples. 
}
\scriptsize
\begin{ruledtabular}
\begin{tabular}[c]{lllll} 
Event    & Systematic Error        &  \multicolumn{3}{c}{{ $E_{\nu}$[GeV] }} \\ 
Sample& Component                & 0.2-3.0 &  0.2-0.9  & 0.9-3.0            \\ \hline
$\nu_{\mu}$   & Flux  [\%]           & 6.9       &    7.2       &  9.0                          \\
                        & Cross-sec. [\%] & 15.7     &   15.9     &  16.2                         \\
                        & Detector [\%]    & 3.2        &   3.6        & 4.6                          \\ 
                       & Total [\%]           & 17.5      &   17.8     &  19.1                       \\ \hline
$\nu_e$         & Flux [\%]            & 9.8       &   8.5         &   11.9                       \\
                        & Cross-sec. [\%]& 14.6     &  14.2       &    15.6                    \\ 
                        & Detector [\%]    & 8.5       &    10.0      &    8.9                     \\  
                        & Total  [\%]         & 19.5     &   19.5       &  21.8                   \\
\end{tabular}
\end{ruledtabular}
\label{table:syst_err}
\end{table}
The agreement between data and the prediction of the neutrino flux from $\pi/K$ parents 
indicates that the NuMI beam modeling provides a good description of the observed off-axis $\nu_\mu$ 
flux in MiniBooNE.

A measurement of the target kaon flux is performed using the $\nu_{\mu}$ CCQE sample.
For this purpose the MC prediction is divided into four templates: 
neutrinos from kaons produced in the NuMI target (shown in Fig.~\ref{fig:prl_fig2}), neutrinos from pions produced
in the NuMI target (also shown in Fig.~\ref{fig:prl_fig2}), neutrinos from kaons
produced in non-target materials (i.e. downstream), and  
neutrinos from pions produced in non-target materials.
Neutrinos from target kaons dominate the event rate in the $E_{\nu}>1.2$ GeV region.
The predicted target kaon fraction in this energy range is 83\% of the selected sample.
In order to get a clean measurement of the target kaon contribution,
events with neutrino energies above 1.2 GeV are selected
and a MC template fit to the data is performed.
In the fit, the predicted target kaon component includes uncertainties associated with 
neutrino cross-section model and detector modeling; other MC components 
include these sources of error as well as flux uncertainties.
After fitting the target kaon component of the MC prediction to the data, with the other MC components fixed to their initial values,
the fit yields a kaon flux that is 1.14 +/- 0.22 compared to the initial MC prediction. The fit has a $\chi^2$/NDF = 9.8/10. 
Therefore, the measured flux of kaons from the NuMI target is consistent with 
the prediction derived from the {\sc FLUKA} cascade model~\cite{fluka}. 
This measurement is a valuable as a check of kaon production which can provide a background in searches 
for $\nu_{\mu} \rightarrow \nu_{e}$  transition in the NuMI beam.

\begin{figure}[ptb]
\includegraphics[height=5.2cm, width=8.6cm]{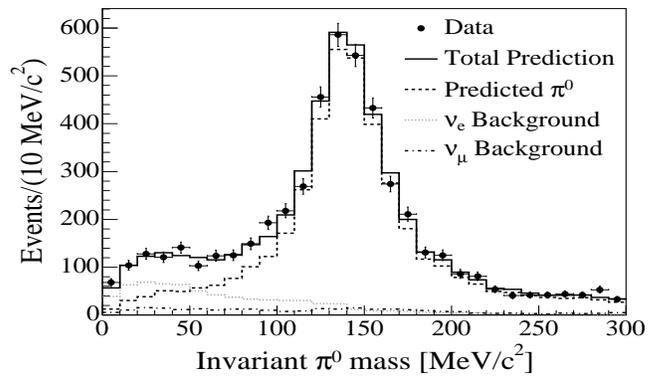}
\caption{Mass distribution of NC $\pi^{0}$ candidates for data (points) and MC (solid histogram).
The dashed histogram is the subset of predicted events with at least one true $\pi^{0}$. 
Predicted non-$\pi^{0}$ backgrounds are either from $\nu_{\mu}$ and $\bar{\nu}_{\mu}$ (dash-dotted line) or $\nu_{e}$ and $\bar{\nu}_{e}$ (dotted line) interactions. Kaon parents contribute 84\% of the events in this sample.}
\label{fig:prl_fig3}
\end{figure}

The $\nu_e$ CCQE events consist of a single subevent of PMT hits 
($\nu_{e}+n\rightarrow e^{-}+p$).
In 8\% of $\nu_{\mu}$ CCQE events the $\mu^{-}$ is captured on carbon,
resulting in a single subevent.
These events are removed with an energy dependent requirement on the likelihood ratio, 
log($L_{e}/L_{\mu}$). 
The majority of the remaining background
is NC $\pi^{0}$ events with only a single reconstructed electromagnetic track
that mimics a $\nu_{e}$ CCQE event. To test our NC  $\pi^{0}$ prediction, a clean sample of NC $\pi^{0}$ events is reconstructed, as shown in Fig.~\ref{fig:prl_fig3}. 
This sample demonstrates good agreement between data and MC. 
About 30\% of the events in the NC  $\pi^{0}$ sample originate from parents produced in non-target materials.
The majority of $\pi^{0}$ events in the $\nu_{e}$ CCQE sample are rejected by requirements on the reconstructed $\pi^{0}$
mass and the electron to pion likelihood ratio,
applied as a function of visible energy. 

\begin{figure}
\includegraphics[height=5.2cm, width=8.6cm]{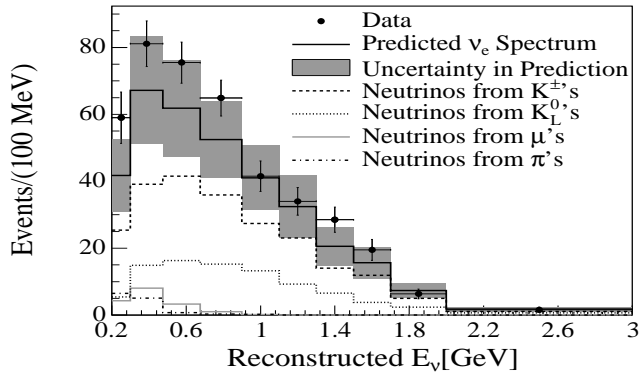} 
\caption{ Reconstructed $E_{\nu}$ distribution of $\nu_{e}$ CCQE candidates. 
The prediction is separated into contributions from neutrino parents.
The band indicates the total systematic uncertainty associated with the MC prediction. 
Kaon parents contribute 93\% of the events in this sample.}
\label{fig:prl_fig4}
\end{figure}

A source of low energy $\nu_{e}$ events arises from the decay of stopped
kaons in the beam stop at the end of the NuMI beamline, which is under the
MiniBooNE detector (see Fig.~\ref{fig:prl_fig1}), 83~m from its center.
Given the kinematics of stopped kaon decay, all $\nu_{e}$'s from this source will have visible
energies ($E_{e}$) below 200 MeV. A requirement 
$E_{e}>200$~MeV effectively removes this source. 
A total of 780 data events pass all of the $\nu_{e}$ CCQE selection criteria.
The MC prediction is 660$\pm$129 
with a  $\nu_{e}$ CCQE efficiency of 32\% and purity of 70\%. 
About 38\% of the events in the $\nu_{e}$ CCQE sample originate from parents produced in non-target materials.
The corresponding energy distribution is shown in Fig.~\ref{fig:prl_fig4}, and
 the uncertainties on the predicted event rate are given in Table~\ref{table:syst_err}.
To facilitate further comparison, the low and high energy regions 
are divided at 0.9~GeV, and 
the numbers of data and MC events in these two regions 
provided in Table~\ref{table:two_bins}.
The data with $E_{\nu}<0.9$~GeV are systematically above the prediction at the 1.2 $\sigma$ level.
\begin{table}
\caption{ Observed and predicted $\nu_e$-like events in two energy bins. The
predicted events are further separated into intrinsic $\nu_{e}$ (and $\bar{\nu}_{e}$) 
and $\nu_{\mu}$ (and $\bar{\nu}_{\mu}$) components.}
\scriptsize
\begin{ruledtabular}
\begin{tabular}[c]{ccccc} 
$E_{\nu}$[GeV] & Data                & MC Prediction     & $\nu_{e}$+$\bar{\nu}_{e}$ & $\nu_{\mu}$+$\bar{\nu}_{\mu}$  \\ \hline
0.2-0.9                 & 496$\pm$22 & 401$\pm$78 & 311                            & 90                                  \\
0.9-3.0                 & 284$\pm$17 & 259$\pm$56 & 231                            & 28                                  \\ 
\end{tabular}
\end{ruledtabular}
\label{table:two_bins}
\end{table}

The results of the measurements described here show that reliable predictions for an off-axis beam 
can be made. However, it should be noted that the MiniBooNE experiment was not constructed as an off-axis detector for NuMI, 
but rather an on-axis detector for the BNB. The uncertainties in the neutrino flux presented here are 
substantially higher than might be expected in a long-baseline off-axis neutrino experiment, due to 
the close proximity of the MiniBooNE detector to the decay pipe and beam dump and the hadronic 
interactions therein which produce neutrinos. Consideration of such items will be important for future experiments being
proposed or executed~\cite{T2K,Nova,Dusel}.

In summary, we have presented the first observation and analysis of neutrino interactions with
an off-axis horn-focused neutrino beam. 
The agreement between data and prediction in the $\nu_{\mu}$ and $\nu_{e}$ CCQE samples demonstrates good 
understanding of both pion and kaon contributions to the beam. 
This represents a successful demonstration of an off-axis neutrino beam at 110~mrad and provides a clear 
proof-of-principle of the off-axis beam concept planned for use in future neutrino experiments. 

\begin{acknowledgments}
We acknowledge the support of Fermilab, the Department of Energy, and
the National Science Foundation. 
\end{acknowledgments}
\end{document}

%% file: collaborators_v2.tex
\newcommand{\UA}{University of Alabama; Tuscaloosa, AL 35487}
\newcommand{\BNL}{Brookhaven National Laboratory; Upton, NY 11973}
\newcommand{\Buchnell}{Bucknell University; Lewisburg, PA 17837}
\newcommand{\Cincinnati}{University of Cincinnati; Cincinnati, OH 45221}
\newcommand{\Colorado}{University of Colorado; Boulder, CO 80309}
\newcommand{\Columbia}{Columbia University; New York, NY 10027}
\newcommand{\EmbryRiddle}{Embry-Riddle Aeronautical University; Prescott, AZ 86301}
\newcommand{\FNAL}{Fermi National Accelerator Laboratory; Batavia, IL 60510}
\newcommand{\Florida}{University of Florida; Gainesville, FL 32611}
\newcommand{\Indiana}{Indiana University; Bloomington, IN 47405}
\newcommand{\LANL}{Los Alamos National Laboratory; Los Alamos, NM 87545}
\newcommand{\LSU}{Louisiana State University; Baton Rouge, LA 70803}
\newcommand{\Michigan}{University of Michigan; Ann Arbor, MI 48109}
\newcommand{\MIT}{Massachusetts Institute of Technology; Cambridge, MA 02139}
\newcommand{\Princeton}{Princeton University; Princeton, NJ 08544}
\newcommand{\SaintMary}{Saint Mary's University of Minnesota; Winona, MN 55987}
\newcommand{\UTAustin}{University of Texas; Austin, TX 78712}
\newcommand{\Tufts}{Tufts University; Medford, MA 02155}
\newcommand{\Virginia}{Virginia Polytechnic Institute \& State University; Blacksburg, VA 24061}
\newcommand{\WIU}{Western Illinois University; Macomb, IL 61455}
\newcommand{\CWM}{College of William \& Mary; Williamsburg, VA 23187}
\newcommand{\Yale}{Yale University; New Haven, CT 06520}

\affiliation{\UA}
\affiliation{\BNL}
\affiliation{\Buchnell}
\affiliation{\Cincinnati}
\affiliation{\Colorado}
\affiliation{\Columbia}
\affiliation{\EmbryRiddle}
\affiliation{\FNAL}
\affiliation{\Florida}
\affiliation{\Indiana}
\affiliation{\LANL}
\affiliation{\LSU}
\affiliation{\Michigan}
\affiliation{\MIT}
\affiliation{\Princeton}
\affiliation{\SaintMary}
\affiliation{\UTAustin}
\affiliation{\Tufts}
\affiliation{\Virginia}
\affiliation{\WIU}
\affiliation{\CWM}
\affiliation{\Yale}


\author{P.~Adamson}\affiliation{\FNAL}
\author{A.~A.~Aguilar-Arevalo}\altaffiliation{Present address: Instituto de Ciencias Nucleares, UNAM, D.F., Mexico}\affiliation{\Columbia}
\author{C.~E.~Anderson}\affiliation{\Yale} 
\author{A.~O.~Bazarko}\affiliation{\Princeton}
\author{M.~Bishai}\affiliation{\BNL}
\author{S.~J.~Brice}\affiliation{\FNAL}
\author{B.~C.~Brown}\affiliation{\FNAL}
\author{L.~Bugel}\affiliation{\Columbia}
\author{J.~Cao}\affiliation{\Michigan}
\author{B.~C.~Choudhary}\affiliation{\FNAL}
\author{L.~Coney}\affiliation{\Columbia}
\author{J.~M.~Conrad}\affiliation{\MIT}
\author{D.~C.~Cox}\affiliation{\Indiana} 
\author{A.~Curioni}\affiliation{\Yale}
\author{Z.~Djurcic}\affiliation{\Columbia}
\author{D.~A.~Finley}\affiliation{\FNAL}
\author{B.~T.~Fleming}\affiliation{\Yale}
\author{R.~Ford}\affiliation{\FNAL}
\author{H.~R.~Gallagher}\affiliation{\Tufts}
\author{F.~G.~Garcia}\affiliation{\FNAL}
\author{G.~T.~Garvey}\affiliation{\LANL}
\author{C.~Green}\affiliation{\LANL}\affiliation{\FNAL}
\author{J.~A.~Green}\affiliation{\Indiana}\affiliation{\LANL}
\author{D.~Harris}\affiliation{\FNAL}
\author{T.~L.~Hart}\affiliation{\Colorado}
\author{E.~Hawker}\affiliation{\WIU}
\author{J.~Hylen}\affiliation{\FNAL}
\author{R.~Imlay}\affiliation{\LSU}
\author{R.~A.~Johnson}\affiliation{\Cincinnati}
\author{G.~Karagiorgi}\affiliation{\MIT}
\author{P.~Kasper}\affiliation{\FNAL}
\author{T.~Katori}\affiliation{\Indiana} 
\author{T.~Kobilarcik}\affiliation{\FNAL}
\author{S.~Kopp}\affiliation{\UTAustin}
\author{I.~Kourbanis}\affiliation{\FNAL}
\author{S.~Koutsoliotas}\affiliation{\Buchnell}
\author{E.~M.~Laird}\affiliation{\Princeton}
\author{S.~K.~Linden}\affiliation{\Yale}
\author{J.~M.~Link}\affiliation{\Virginia}
\author{Y.~Liu}\affiliation{\Michigan} 
\author{Y.~Liu}\affiliation{\UA}
\author{L.~Loiacono}\affiliation{\UTAustin}
\author{W.~C.~Louis}\affiliation{\LANL}
\author{A.~Marchionni}\affiliation{\FNAL}
\author{K.~B.~M.~Mahn}\affiliation{\Columbia}
\author{W.~Marsh}\affiliation{\FNAL}
\author{G.~McGregor}\affiliation{\LANL}
\author{M.~D.~Messier}\affiliation{\Indiana}
\author{W.~Metcalf}\affiliation{\LSU}
\author{P.~D.~Meyers}\affiliation{\Princeton}
\author{F.~Mills}\affiliation{\FNAL}
\author{G.~B.~Mills}\affiliation{\LANL}
\author{J.~Monroe}\affiliation{\MIT}
\author{C.~D.~Moore}\affiliation{\FNAL}
\author{J.~K.~Nelson}\affiliation{\CWM}
\author{R.~H.~Nelson}\affiliation{\Colorado}
\author{V.~T.~Nguyen}\affiliation{\MIT}
\author{P.~Nienaber}\affiliation{\SaintMary}
\author{J.~A.~Nowak}\affiliation{\LSU}
\author{S.~Ouedraogo}\affiliation{\LSU}
\author{R.~B.~Patterson}\affiliation{\Princeton}
\author{Z.~Pavlovic}\affiliation{\UTAustin}
\author{D.~Perevalov}\affiliation{\UA}
\author{C.~C.~Polly}\affiliation{\Indiana}
\author{E.~Prebys}\affiliation{\FNAL}
\author{J.~L.~Raaf}\affiliation{\Cincinnati}
\author{H.~Ray}\affiliation{\LANL}\affiliation{\Florida}
\author{B.~P.~Roe}\affiliation{\Michigan}
\author{A.~D.~Russell}\affiliation{\FNAL}
\author{V.~Sandberg}\affiliation{\LANL}
\author{R.~Schirato}\affiliation{\LANL}
\author{D.~Schmitz}\affiliation{\Columbia}
\author{M.~H.~Shaevitz}\affiliation{\Columbia}
\author{F.~C.~Shoemaker}\affiliation{\Princeton}
\author{W.~Smart}\affiliation{\FNAL}
\author{D.~Smith}\affiliation{\EmbryRiddle}
\author{M.~Sodeberg}\affiliation{\Yale} 
\author{M.~Sorel}\affiliation{\Columbia}
\author{P.~Spentzouris}\affiliation{\FNAL}
\author{I.~Stancu}\affiliation{\UA}
\author{R.~J.~Stefanski}\affiliation{\FNAL} 
\author{M.~Sung}\affiliation{\LSU} 
\author{H.~A.~Tanaka}\affiliation{\Princeton}     
\author{R.~Tayloe}\affiliation{\Indiana}
\author{M.~Tzanov}\affiliation{\Colorado}
\author{P.~Vahle}\affiliation{\CWM}
\author{R.~Van~de~Water}\affiliation{\LANL}
\author{B.~Viren}\affiliation{\BNL}
\author{M.~O.~Wascko}\altaffiliation{Present address: Imperial College London, London, UK}\affiliation{\LSU}
\author{D.~H.~White}\affiliation{\LANL}
\author{M.~J.~Wilking}\affiliation{\Colorado}
\author{H.~J.~Yang}\affiliation{\Michigan}
\author{F.~X.~Yumiceva}\affiliation{\CWM}
\author{G.~P.~Zeller}\affiliation{\Columbia}\affiliation{\LANL}
\author{E.~D.~Zimmerman}\affiliation{\Colorado}
\author{R.~Zwaska}\affiliation{\FNAL}